\def\sqr#1#2{{\vcenter{\vbox{\hrule height.#2pt\hbox{\vrule
width.#2pt height#1pt \kern#1pt\vrule width.#2pt}\hrule height.#2pt}}}}

\documentclass[12pt,a4paper]{article} 

\font\cst=cmr10 scaled \magstep4
\font\csc=cmr10 scaled \magstep2

\begin{document}

\vglue 2cm

\centerline{\cst Conserved Charges}
\vskip 0.7cm
\centerline{\cst in Einstein Gauss-Bonnet theory}
\vskip 2cm

\centerline{\bf Nathalie Deruelle} 
\vskip 0.2 cm
\centerline{\it  Institut d'Astrophysique de Paris,}
\centerline{\it GReCO, FRE 2435 du CNRS,}
\centerline{\it 98 bis Boulevard  Arago, 75014 Paris, France}
\vskip 0.7 cm
\centerline{\bf Joseph Katz}
\vskip 0.2cm
\centerline{\it Racah Institute of Physics, Hebrew University}
\centerline{\it 91904, Jerusalem, Israel}
\vskip 0.7 cm
\centerline{\bf Sachiko Ogushi}
\vskip 0.2cm
\centerline{\it Yukawa Institute for Theoretical Physics}
\centerline{\it Kyoto University, Kyoto 606-8502, Japan}

\vskip 1.5cm
\centerline{\today}

\vskip 2cm
\noindent
{\bf Abstract}

Using Noether's identities, we define a superpotential with respect to a background for the Einstein Gauss-Bonnet  theory of gravity. As an example, we show that its associated
conserved charge  yields the mass-energy of a D-dimensional Gauss-Bonnet black hole in an anti-de Sitter spacetime.

\vfill\eject

\bigskip
\noindent
{\csc 0. Introduction}
\medskip
Einstein Gauss-Bonnet gravity has received renewed interest recently,
both in the context of the Randall-Sundrum  and braneworld scenarios
and in the context of the AdS/CFT
correspondence, see e.g.~\cite{DerMad03} for a recent review. In particular,
well-known black hole solutions~\cite{BouDes85} and their recent generalizations
~\cite{Cai01},  have been extensively reconsidered, with particular emphasis on
the various ways of calculating their entropy~\cite{MyeSim88}.

In the widely used Euclidean approach~\cite{GibHaw77}~\cite{HawPag83} 
the now standard steps to get that entropy are :
start from, say, an exact static black hole solution depending on the
integration constant $\mu$ (the ``mass parameter" ); compute, as a
function of $\mu$,  the Euclidean action $I$ on shell;  then obtain the
``internal" energy as $E={\partial I\over\partial\beta}$ and, finally,
the entropy as $S=\beta E-I$, where $\beta$ is the  period in Euclidean
time which regularizes the action on shell. This method has been recently
successfully re-applied, following~\cite{MyeSim88}, to anti-de Sitter (AdS)
Gauss-Bonnet black holes, see
e.g.~\cite{Cai01,NojOdiOgu01,CveNojOdi01,ChoNeu02,Neu02,Neu03}. 
In particular, it relates in a definite manner the
internal energy $E$ to the mass parameter $\mu$.

Now, the mass parameter $\mu$ also appears in the coefficient of the
$1/R^{D-3}$ term in the asymptotic form of the metric and is related to
the ``gravitational" mass-energy $M$ of the black hole. As expected,
the internal energy  $E$ turns out to be nothing than $M$. That result
though is by no means trivial. Indeed, the Euclidean method applies to
quantum radiating black holes whereas the notion of the gravitational
mass of a spacetime is a purely classical one, related to the motion of
test particles for example, and not based on the existence of a horizon
and its temperature. Moreover, there is yet another way to define the
mass of a spacetime, to wit by means of conserved quantities. In pure
Einstein theory, this alternative definition has been shown to coincide
with the previous ones : for an introduction and references to the many
different ways to define, in a purely classical setup, conserved
quantities associated with a spacetime, see e.g.~\cite{BroLauYor00}.
In the present paper we extend to Einstein Gauss-Bonnet theory of
gravity a definition of conserved quantities with respect to  a
background which uses a superpotential and
(asymptotically) Killing vectors. This approach has been advocated by
Rosen~\cite{Ros63} and developped in ~\cite{Kat85,KatBicLyn97}. 
When applied to AdS Gauss-Bonnet black holes it will yield the sought 
for equality  of their ``conserved", ``internal" and ``gravitational" energies. 
Moreover, the calculation will be straightforward (as compared to the Euclidean
method) and hence provide a short cut for calculating the entropy.

Other methods based on superpotentials have been proposed. For example, Deser and Tekin~\cite{DesTek02} extend to Einstein Gauss-Bonnet theory the definition of conserved
energy as proposed by Abbot-Deser~\cite{AbbDes82} and apply it to AdS Gauss-Bonnet black holes (see~\cite{Pad03} for a generalization to arbitrary backgrounds using a
Hamiltonian approach). It must be noted however that, contrarily to the  ``KBL" superpotential~\cite{KatBicLyn97}, the AD formula~\cite{AbbDes82} does not yield the Bondi
mass, see~\cite{PetKat02}.  As for Hamiltonian methods, which are particularly suited to obtaining the energy, see e.g.~\cite{Wal93}, they have been extended to Einstein
Gauss-Bonnet theory in~\cite{JacMye93}~\cite{JacKanMye95}~\cite{BanTeiZan93}~\cite{LouSimWin96} (for a  covariant formalism, see~\cite{AllFraRai03}).

The paper is organized as follows : in Section I we recall the standard procedure to obtain Noether's conserved currents from a Lagrangian as initiated in the 50's 
(see, for instance~\cite{BerSch53}) and we summarize the formalism developped  in~\cite{Kat85}~\cite{KatBicLyn97}  to define conserved currents and charges of a given
spacetime with respect to a background. 

In Section II we specialize to Einstein Hilbert's Lagrangian~: we recall the expressions obtained in~\cite{Kat85}~\cite{KatBicLyn97} for the superpotential and 
we compute, as a warm up exercise, the associated  conserved charges in two simple cases~:  the mass of a D-dimensional, static, spherically symmetric and  asymptotically
flat or anti-de Sitter spacetime on one hand and the angular momentum of the BTZ solution~\cite{BanTeiZan92} on the other. 

In Section III, which is the core of the paper, we extend the previous formalism and results to the Gauss-Bonnet Lagrangian and, as an application, we compute the mass of
Einstein Gauss-Bonnet D-dimensional black holes.

Finally in Section IV we examine some of the shortcomings of our approach.

\bigskip
\noindent
{\csc I. The general setting~: a short review}
\medskip

{\bf a. Construction of superpotentials and definition of conserved charges from a Lagrangian}
\medskip

Consider, in a $D$-dimensional spacetime with coordinate system $X^A$, metric coefficients $g_{AB}$  (mostly plus signature) and associated covariant derivative $D_A$, a
family
$${\cal L}=L+D_Ak^A\eqno(1.1)$$
 of Lagrangians for pure gravity. $L$ is a scalar quantity (some function of curvature invariants, functional of the metric, its first and second derivatives) and $k^A$ are the $D$
components of a vector which enters the boundary term in the variation of the action (see below). A hat will denote multiplication by $\sqrt{-g}$ where $g$ is the
determinant of  $g_{AB}$. The variation of the scalar  density $\hat{\cal L}$ with respect to the metric can be written formally as
$$\delta\hat{\cal L}=-\hat\sigma^{AB}\delta g_{AB}+\partial_A(\hat V^A+\delta \hat k^A)\,. \eqno(1.2)$$
 The tensor $\sigma^{AB}$ is the variational derivative of $\hat{\cal L}$ and may  contain derivatives of the metric up to  fourth order. The field equations which govern pure
gravity are $\sigma^{AB}=0$, whatever the vector $k^A$. As for the vector $V^A$ it can be written under the form
$$V^A=\alpha^{A(BC)}\,\delta g_{BC}+\beta^{A(BC)}_{\hskip 0.8cm D}\ \delta\Gamma^D_{BC}\eqno(1.3)$$ 
where $\Gamma^D_{BC}$ are the Christoffel symbols and
$$\delta\Gamma^D_{BC}={1\over 2}g^{DE}(D_B\delta g_{CE}+D_C\delta g_{BE}-D_E\delta g_{BC})\,.$$
$\alpha^{A(BC)}$ and $\beta^{A(BC)}_{\hskip 0.8cm D}$ are tensors which are third and second order in the metric derivatives respectively. Parentheses denote symmetrisation
as in $\alpha^{A(BC)}={1\over2}(\alpha^{ABC}+\alpha^{ACB})$.

If the variation $\delta g_{AB}$ is due to a mere change of coordinates $X^A\to\tilde X^A=X^A-\xi^A$ with $\xi^A$ an infinitesimal vector, then $\delta$ reduces to a Lie
derivative and 
$$\delta g_{AB}=2D_{(A}\xi_{B)}\qquad,\qquad\delta\Gamma^D_{BC}=D_{(BC)}\xi^D-R^D_{\ \ (BC)E}\,\xi^E\eqno(1.4)$$
 where $R^D_{\ \ BCE}=\partial_C\Gamma^D_{BE}+...$ is the Riemann tensor. The variation (1.2) of the Lagrangian density  can in that case be cast in the following form~:
$$\partial_A\left[(2\hat\sigma^{AB}+g^{AB}\hat L)\xi_B-\hat V^A+2D_B(\xi^{[A}\hat k^{B]})\right]=2\hat\xi^BD_A\sigma^{AB}\,.\eqno(1.5)$$
Antisymmetrisation is denoted by  brackets as in $\xi^{[A}\hat k^{B]}={1\over 2}(\xi^A\hat k^B-\xi^B\hat k^A)$. Now the right hand side of (1.5) is identically zero. This is
immediately clear from Stokes' theorem if we  integrate (1.5) over an arbitrary domain of spacetime  with any vector $\xi^A$ which is  zero on the boundary. In Einstein's or 
Einstein Gauss-Bonnet's theory $D_A\sigma^{AB}=0$ are the (generalized) Bianchi identities.

Rearranging $V^A$, we then have that the ``current"
$$j^A=j^A_a+j^A_b\eqno(1.6)$$ 
with
$$ j^A_a=\left(Lg^{AB}+2\sigma^{AB}+\beta^{A(CD)E}\,R^B_{\ CDE}\right)\xi_B-2\alpha^{A(BC)}\,D_B\xi_C-\beta^{A(BC)D}D_{BC}\xi_D\eqno(1.7)$$
and
$$j^A_b=2D_B(\xi^{[A} k^{B]})\eqno(1.8)$$
 is  {\it identically  } conserved~:
$$D_Aj^A\equiv 0\qquad\Longleftrightarrow\qquad\partial_A\hat j^A \equiv 0\,.\eqno(1.9)$$

The conservation of $j^A$ implies that there exists an antisymmetric ``superpotential" $j^{[AB]}$ such that 
$$\hat j^A\equiv D_B\hat j^{[AB]}=\partial_B\hat j^{[AB]}\,.\eqno(1.10)$$
 Looking for an expression of the form
$$j^{[AB]}=j^{[AB]}_a+j^{[AB]}_b\eqno(1.11)$$
 with
$$ j^{[AB]}_a={\cal A}^{[AB]C}\xi_C+{\cal B}^{[AB]CD}D_C\xi_D\qquad{\rm and}\qquad j^{[AB]}_b=2\xi^{[A} k^{B]}\eqno(1.12)$$ 
we get from  (1.7)
$${\cal A}^{[AB]C}=-2\alpha^{A(BC)}-D_D\,{\cal B}^{[AD]BC}$$
$${\cal B}^{[AB]CD}+{\cal B}^{[AC]BD}=-2\beta^{A(BC)D}\eqno(1.13)$$
$$D_B{\cal A}^{[AB]E}-{1\over2}{\cal B}^{[AB]CD}\ R^E_{\ DBC}=Lg^{AE}+2\sigma^{AE}+\beta^{A(CD)B}\,R^E_{\ CDB}\,.$$ 
${\cal B}^{[AB]CD}$ can be extracted from the second equation (1.13), the first one yields ${\cal A}^{[AB]C}$ and the third is an identity (which can serve to check the solution
found for ${\cal B}^{[AB]CD}$ and  ${\cal A}^{[AB]C}$).

Consider now a region ${\cal M}$ of spacetime where (1.9) applies, bounded by $\partial{\cal M}$, a timelike hypercylinder limited by 2 spacelike hypersurfaces  $X^0=t_0$
and $X^0=t_1$, and apply Stokes' theorem (N.B.~: if $j^A$ is singular at the origin a  volume around it may have to be excluded)~: 
$$\int_{\cal M}d^Dx\,\partial_A\hat j^A=0\qquad\Longleftrightarrow\qquad\int_{\partial{\cal M}}\!d^{D-1}x\ n_\mu\hat j^\mu=0$$
$$\Longleftrightarrow\quad\int_{t=t_1}\!d^{D-1}x\ \hat j^0-\int_{t=t_0}\!d^{D-1}x\ \hat j^0+\int_{t_0}^{t_1} dt\int_{\cal S}\!d^{D-2}x\ n_i\hat j^i=0\,.\eqno(1.14)$$ 
$d^Dx=(dX^0\times ...dX^{(D-1)})$~;  ${\cal S}$ is the $(D-2)$-dimensional sphere at infinity, with $n^i$ its unit normal vector pointing towards infinity. Using  (1.10), we have
that $\hat j^0=\partial_i\hat j^{[0i]}$, so that the ``charge"
$$q=\int_{\cal S}\!d^{D-2}x\ n_i\hat j^{[0i]}\eqno(1.15)$$
 is independent of time if the last term in (1.14) vanishes. The vanishing of that last term depends on the properties of spacetime (such as stationarity) as well as on the choice
for the vectors $\xi^A$ and $k^A$. (Of course, when ${\cal S}$ is the $(D-2)$-dimensional sphere at infinity, $q$ may not be finite.)
 
\bigskip {\bf b. Choosing the vector $\xi^A$}
\medskip

We are interested in defining conserved charges such as the energy and angular momentum of a given spacetime. Such quantities are related to isometries under time
translations and rotations and hence to the existence of Killing vectors. In the following we shall in fact restrict our attention  to stationary  spacetimes and choose $\xi^A$ to
be a Killing vector~:
$$D_{(A}\xi_{B)}=0\,.\eqno(1.16)$$
 (Note that $\xi^A$ is defined up to an arbitrary constant, see Section IV.)

\bigskip {\bf c. Choosing the vector $k^A$ }
\medskip

 The variation with respect to the metric of the action for pure gravity
$$I=\int_{\cal M}\!d^Dx\ \hat{\cal L}=\int_{\cal M}\!d^Dx\ \hat L+\int_{\partial{\cal M}}\!d^{D-1}x\ n_\mu \hat k^\mu \eqno(1.17)$$
follows from (1.2) and is
$$\delta I=-\int_{\cal M}d^Dx\,\hat\sigma^{AB}\delta g_{AB}+\int_{\partial{\cal M}}\!d^{D-1}x\ n_\mu (\hat V^\mu+\delta\hat k^\mu)\,.\eqno(1.18)$$
 It is customary (see however Section IV) to choose the vector $k^A$ in such a way that the boundary term does not contain terms proportional to normal derivatives of the 
metric variations. In that case indeed, the field equations are obtained by keeping only the metric and its tangential derivatives fixed at the boundary, leaving their normal
derivatives free, and the variational problem falls within the scope of ordinary Lagrangian  field theory.

In view of the expression (1.3) for $V^A$, the vector $k^A$ must be a linear combination of the metric and its first order derivatives, or, equivalently, the Christoffel symbols.
However, since the Christoffel symbols are not tensors, $k^A$ cannot be constructed from the metric and its first derivatives alone. One needs some extra element to get a
covariant formalism, like tetrads~\cite{Mol61}, a  foliation useful in the Hamiltonian formalism~\cite{Yor86} or, as advocated by Rosen~\cite{Ros63} and implemented
in~\cite{Kat85}, a background. We shall define our conserved quantities with respect to a background. This method is well suited to the applications we shall study.

\bigskip {\bf d. Superpotential and charge with respect to a background}
\medskip

More precisely we shall consider two spacetimes, ${\cal M}$ and $\bar{\cal M}$, with respective metrics and covariant metric derivatives ($g_{AB}, D_C$) and ($\bar g_{AB},
\bar D_C$), together with a mapping which associates a point $P$ in ${\cal M}$ with coordinates $X^A$ to a point $\bar P$ in $\bar{\cal M}$ with the same coordinates $X^A$.
Then
$$\Delta^A_{BC}\equiv\Gamma^A_{BC}-\bar\Gamma^A_{BC}={1\over2}g^{AD}(\bar D_Bg_{CD}+\bar D_Cg_{DB}-\bar D_Dg_{BC})\eqno(1.19)$$
 is a tensor under coordinate transformations which are the same in both spacetimes. (In the following a bar will denote a quantity constructed out of the metric $\bar g_{AB}$
and its associated covariant metric derivative $\bar D_A$.) The action is now defined by $I-\bar I$, the Lagrangian density by $\hat{\cal
L}-\bar{\hat {\cal L}}$ and the corresponding superpotential density  becomes
$$\hat J^{[AB]}={1\over2\kappa}\left(\hat j^{[AB]}-\overline{\hat j^{[AB]}}\right)\,,\eqno(1.20)$$
where $\kappa$ is a coupling constant. This is the way it was defined in~\cite{KatBicLyn97} and it is sometimes referred to as the KBL superpotential. 

The corresponding charge $Q$ of ${\cal M}$ with respect to $\bar{\cal M}$ is defined similarly as 
$$Q=\int_{\cal S}\!d^{D-2}x\ n_i\hat J^{[0i]}\,.\eqno(1.21)$$

\bigskip
\noindent
{\csc II. The example of Einstein's theory}
\medskip

In Einstein's theory, pure gravity is described by the Einstein-Hilbert Lagrangian
$$L_{[1]}=-2\Lambda+s\,.\eqno(2.1)$$
 $\Lambda$ accounts for a possible cosmological constant and  $s$ is the scalar curvature of spacetime~: $s=g^{AB}r_{AB}$, $r_{AB}$ being the Ricci tensor~:
$r_{AB}=R^C_{\ ACB}$. The variation of (1.1) with respect to the metric then gives (1.2) with
$$\sigma^{[1]}_{AB}=\Lambda g_{AB}+\sigma^{(1)}_{AB}\quad,\quad \sigma^{(1)}_{AB}\equiv r_{AB}-{1\over2}\,s\, g_{AB}\quad;\quad
V_{[1]}^A=-\left(g^{AB}\delta\Gamma^C_{BC}-g^{BC}\delta\Gamma^A_{BC}\right)\,,\eqno(2.2)$$
hence, see (1.3), $\alpha_{[1]}^{A(BC)}=0$ and $\beta_{[1]}^{A(BC)D}=g^{AD}g^{BC}-g^{A(B}g^{C)D}$.
Solving equations (1.13) readily yields ${\cal B}_{[1]}^{[AB]CD}=2g^{C[A}g^{B]D}$ 
and ${\cal A}_{[1]}^{[AB]C}=0$, so that the superpotential (1.12)  is given by
$$j_{[1]}^{[AB]}=j_{[1]a}^{[AB]}+j_{[1]b}^{[AB]} \qquad\hbox{with}\qquad j_{[1]a}^{[AB]} =2D^{[A}\xi^{B]}\quad\hbox{and}\quad j_{[1]b}^{[AB]} =2\xi^{[A}k_{[1]}^{B]}\,.\eqno(2.3)$$
 As for the associated charge (1.15) it becomes
$$q_{[1]}=q_{[1]a}+q_{[1]b}\quad\hbox{with}\quad q_{[1]a}=2\int_{\cal S}\!d^{D-2}x\ n_i\,\hat D^{[0}\xi^{i]}\quad\hbox{and}\quad q_{[1]b}=2\int_{\cal S}\!d^{D-2}x\
n_i\,\hat\xi^{[0}k_{[1]}^{i]}\,.
\eqno(2.4)$$
\newpage
\bigskip 
{\bf a. The KBL vector, superpotential and  charge}
\medskip

A straightforward choice for  $k_{[1]}^A$, which cancels all the terms proportional to $\delta\Gamma^D_{BC}$ in $V_{[1]}^A$ and also cancels not only normal but all
 derivatives of the metric variations $\delta g_{AB}$  in the boundary term of $\delta I$, see (1.18), is obtained by replacing $\delta\Gamma^A_{BC}$ by $-\Delta ^A_{BC}$
in the expression (2.2) for $V_{[1]}^A$ and is\footnote{Note that the action (1.17) with $k_{[1]}^A$ defined by (2.5) is {\it not} the Einstein Gibbons-Hawking
action~\cite{GibHaw77}.}
$$k_{[1]}^A=-\beta_{[1]}^{A(BC)D}\Delta^D_{BC}={\cal B}_{[1]}^{[AB]CD}\Delta^D_{BC}=g^{AB}\Delta^C_{BC}-g^{BC}\Delta^A_{BC}\,. \eqno(2.5)$$

 The superpotential density $\hat J_{[1]}^{[AB]}$ is 
$$\hat J_{[1]}^{[AB]}=\hat J_{[1]a}^{[AB]} +\hat J_{[1]b}^{[AB]} \eqno(2.6)$$
 with
$$\kappa\hat J^{[AB]}_{[1]a}=\hat D^{[A}\xi^{B]}-\overline{\hat D^{[A}\xi^{B]}}\qquad\hbox{and}\qquad \kappa \hat J^{[AB]}_{[1]b} =\xi^{[A}\hat k_{[1]}^{B]}\,.\eqno(2.7)$$
The corresponding charge $Q_{[1]}$ of ${\cal M}$ with respect to $\bar{\cal M}$ is
$$Q_{[1]}={1\over\kappa}\int_{\cal S}\!d^{D-2}x\ n_i\left(\hat D^{[0}\xi^{i]}-\overline{\hat D^{[0}\xi^{i]}}+\hat\xi^{[0}k_{[1]}^{i]}\right)\,.\eqno(2.8)$$

Provided adequate normalisation of the Killing vector $\xi^A$ (for example $\xi^A=(1,0,0,0)$ to get the conserved mass, which is a natural choice in asymptotically flat
four dimensional spacetimes as it identifies proper and coordinate times at infinity), and adequate definition of
$\kappa$ ($\kappa=-{8\pi G\over c^4}$ where $G$ is Newton's constant and $c$ the speed of light, which is also the natural choice when one considers coupling to matter
fields), the KBL superpotential (2.6-7) is the {\it only} superpotential which satisfies all the following properties~\cite{JulSil00}~:

$\bullet$ It is generally covariant and then can be computed in any coordinate system,

$\bullet$ If the chosen coordinates are the {\it Cartesian} ones of an asymptotically flat (or AdS) spacetime, the KBL superpotential gives the ADM mass
formula\footnote{references are given in~\cite{JulSil00}} or the AD mass,

$\bullet$ It gives the mass and angular momentum (and the Brown-Henneaux conformal charge for $AdS_3$) with the right normalization in any spacetime of dimension $D\ge
3$. More generally it can be used for any asymptotic Killing vector $\xi^A$.

It has also been proven \cite{KatLer96} that at null infinity, the KBL superpotential yields the Bondi mass, the Penrose linear momentum and the Penrose-Dray-Streubel
angular momentum.

There are thus good reasons to try and find the  generalization of the KBL superpotential to Einstein-Gauss-Bonnet Lagrangians. Before proceeding however we shall, as a warm
up exercise and because the KBL superpotential has rarely been applied to non asymptotically flat backgrounds,  compute its associated conserved charges in two different
examples.

\newpage
\bigskip {\bf b. The KBL mass of a static, spherically symmetric and
asymptotically anti-de Sitter spacetime}
\medskip

Consider first a finite static, spherically symmetric distribution of matter  in a $D$-dimensional  anti-de Sitter background. The metric outside matter can be written as
(setting $X^0\equiv cT$ and $X^1\equiv R$)
$$ds^2=-f\, c^2\,dT^2+{dR^2\over f}+R^2d\Omega^2_{(D-2)}\eqno(2.9)$$
 where $f$ is a function of $R$ and where $d\Omega^2_{(D-2)}$ is the line element of a $(D-2)$-dimensional unit sphere\footnote{for example~: $d\Omega^2_{(1)}=d\phi^2$,
$d\Omega^2_{(2)}=d\theta^2+\sin^2\theta\,d\phi^2$, $d\Omega_{(3)}=d\chi^2+\sin^2\chi (d\theta^2+\sin^2\theta\,d\phi^2)$, etc}. When gravity is described by Einstein's theory
with
$\Lambda<0$,   $f$ is given (outside matter) by the Schwarzschild anti-de Sitter solution
$$f(R)=1-{\mu\over R^{D-3}}+{R^2\over \ell^2}\eqno(2.10)$$
 where $\ell^2=-{(D-1)(D-2)\over2\Lambda}$  and where the integration constant $\mu$ is the ``mass" parameter, that we want to relate to the KBL conserved mass of
Schwarzschild anti-de Sitter spacetime as defined previously.

It is quite natural to choose the background  spacetime $\bar{\cal M}$ to be the $\mu=0$ anti-de Sitter spacetime (or Minkowski if the cosmological constant is set to zero),
that is
$$\bar f=1+{R^2\over \ell^2}\qquad\hbox{so that}\qquad \Delta f\equiv f-\bar f=-{\mu\over R^{D-3}}\,.\eqno(2.11)$$

 In asymptotically flat spacetimes the timelike Killing vector associated with energy is normalized to represent translations at spatial infinity. Here we cannot use such
normalization. It is customary (see e.g.~\cite{DesTek02}), although by no means fully justified (see e.g.~\cite{JacMye93}), to normalize the Killing vector to local $T$
translations of the  AdS background : 
$$\xi^A=(1,0,...,0)\,.\eqno(2.12)$$

The conserved mass $M_{[1]}$ of Schwarzschild anti-de Sitter spacetime (with respect to the  anti-de Sitter  background) is therefore defined as, see (2.8) and denoting
$Q_{[1]}=M_{[1]}\,c^2$~:
$$M_{[1]}=M_{[1]a}+M_{[1]b}\eqno(2.13)$$
 with
$$\kappa M_{[1]a}c^2=\int_{\cal S}\!d^{D-2}x\ \left(\hat D^{[0}\xi^{1]}-\overline{\hat D^{[0}\xi^{1]}}\right)\eqno(2.14)$$
 and
$$ \kappa M_{[1]b}\,c^2=\int_{\cal S}\!d^{D-2}x\ \hat\xi^{[0}k^{1]}_{[1]}\eqno(2.15)$$
 where $\kappa$ is a coupling constant that we shall relate to a $D$-dimensional ``Newton" constant $G_D$ and the speed of light $c$ by
$$\kappa=-{8\pi G_D\over c^4}\,,\eqno(2.16)$$
 so that it reduces to the natural coupling in four dimensions.

The contribution of $M_{[1]a}$ to the mass is obtained readily~: since $D^{[0}\xi^{1]}=-{1\over 2}f'$ (where a prime denotes differentiation with respect to $X^1\equiv R$), we
have, see (2.14)
$$\kappa M_{[1]a}\, c^2=-{\cal V}_{(D-2)}\lim_{R\to\infty}\,R^{D-2}\ {\Delta f'\over2}\eqno(2.17)$$
 where ${\cal V}_{(D-2)}$ is the volume of the unit $(D-2)$-dimensional unit sphere~; hence, from (2.11)
$$\kappa M_{[1]a}\,c^2=-{\cal V}_{(D-2)}\, {(D-3)\over2}\,\mu\,.\eqno(2.18)$$
(In $D=4$ dimensions this is half Komar's value~\cite{Kom59}.)

As for the vector $k^A_{[1]}$ it is given by (2.5) and its $R-$component is easily found to be 
$$k^1_{[1]}=\Delta f'+{(D-2)\over R}\Delta f -{(\Delta f/\bar f)^2\over1+(\Delta f/\bar f)}\,{\bar f'\over2}\,.\eqno(2.19)$$
 If $\bar f=1$ the last term is absent~; if $\bar f=1+R^2/\ell^2$ it is negligible at infinity when $\Delta f$ is given by (2.11). Hence, see (2.15)
$$\kappa M_{[1]b}\, c^2=-{1\over2}\,{\cal V}_{(D-2)}\, \mu\,.\eqno(2.20)$$
(In $D=4$ dimensions this is again half Komar's value~\cite{Kom59}.)

Therefore, using the KBL superpotential, the mass parameter $\mu$ of an asymptotically  Schwarzschild anti-de Sitter spacetime is related to the associated KBL conserved
mass $M_{[1]}=(M_{[1]a}+M_{[1]b})$  by adding (2.17) and (2.20) and is
$$\mu={16\pi G_D\over c^2{\cal V}_{(D-2)}}{M_{[1]}\over (D-2)}\,.\eqno(2.21)$$

Now, in four dimensions, the mass parameter $\mu$ is related to the ``gravitational" mass $M_*$, as deduced from the motion of test particles say, by the
Schwarzschild formula~: $\mu={2GM_*\over c^2}$, a formula which can be generalized to $D$ dimension as
$$\mu={16\pi G_D\over c^2{\cal V}_{(D-2)}}{M_*\over
(D-2)}\,.\eqno(2.22)$$
 Hence, as they should, the gravitational mass $M_*$ and the KBL conserved mass $M_{[1]}$ are equal. In this particular example there is perfect
agreement with~\cite{AbbDes82}.

 If instead of matter around the center there is a black hole, Stokes theorem does no more apply because there is a singularity at the origin and the interpretation of (2.13-15) as
mass-energy becomes problematic because the Killing field is not timelike inside the horizon. It should however be noted that the calculation of the mass of a spherical
spacetime by means of the superpotential and hence as a surface integral at infinity clearly indicates the global topological origin of what we called mass-energy which ignores
the internal geometry of spacetime. The mass-energy is therefore attributed to the whole spacetime whether there is   matter in the center or collapsed matter into a
black-hole.

This being said, Stokes theorem is still applicable to any region of the static spacetime between two spheres. Integration between spatial infinity and the horizon of a black
hole~\cite{BarCarHaw73} combined with Beckenstein's idea~\cite{Bek73} of attributing an entropy to the black hole proportional to its area and the obtention by
Hawking~\cite{Haw75} of their quantum  black body temperature led to a new branch of physics~: black hole thermodynamics.  Knowing the Hawking temperature T and the
angular velocity $\Omega$ of the rotating hole as a function of its conserved mass $M$ and angular momentum $J$, the entropy of the black hole is readily obtained by a simple
quadrature, via the second law~: T$dS=dM -\Omega dJ$.

\vfill
{\bf c. The angular momentum of the BTZ spacetime}
\medskip

As a second example we shall compute the angular momentum of the BTZ solution of the $3$-dimensional Einstein vacuum equations with cosmological constant
$\Lambda$~\cite{BanTeiZan92}. The metric can be written as (denoting $X^0\equiv cT, X^1\equiv R, X^2\equiv\phi$)
$$ds^2=-N^2\, c^2\, dT^2+{dR^2\over N^2}+R^2\left(d\phi-N^\phi c\,dT\right)^2\eqno(2.23)$$
with
$$N^2=-\mu+{R^2\over\ell^2}+{a^2\over R^2}\qquad\hbox{and}\qquad N^\phi=-{a\over R^2}\quad\hbox{(so that}\quad g_{0\phi}=a)\eqno(2.24)$$
where $\ell^2=-\Lambda^{-1}$ and where the integration constants $\mu$ and $a$ are the ``mass" and ``angular momentum" parameters that we want to relate to the KBL
conserved mass and angular momentum as defined previously.

As before we shall choose the background $\bar{\cal M}$ to be the $\mu=0, a=0$ solution for which
$$\bar N^2={R^2\over\ell^2}\qquad\qquad\hbox{and}\qquad\qquad \bar N^\phi=0\,.\eqno(2.25)$$
The two Killing vectors that we shall consider are those associated with time translations and angular rotations and normalized as
$$\xi^A_{(T)}=(1,0,0)\qquad,\qquad\xi^A_{(\phi)}=(0,0,1)\,.\eqno(2.26)$$

A straightforward calculation\footnote{using $-g_{00}\equiv N_1^2=N^2-{a^2\over R^2}$ , $g_{11}={1\over N^2}$ , $g_{\phi\phi}=R^2,g_{0\phi}=a$ as
well as  its determinant
$g=-R^2$ and its inverse $-g^{00}={1\over N^2}$ , $g^{11}=N^2$ , $g^{\phi\phi}={N^2_1\over R^2 N^2}$ , $g^{0\phi}={a\over R^2N^2}$} then yields
$$\hat D^{[0}\xi^{1]}=-\left({R^2\over\ell^2}\xi^0+a\xi^2\right)\eqno(2.27)$$
and hence
$$\hat D^{[0}\xi^{1]}-\overline{\hat D^{[0}\xi^{1]}}=-a\xi^2\,.\eqno(2.28)$$
Another short calculation (see (2.5)) yields
$$\hat\xi^{[0}k^{1]}_{[1]}=\left\{{R^2\over2\ell^2}\left(1-{R^2\over\ell^2N^2}\right)-{a^2\over2R^2}\left(2-{R^2\over\ell^2N^2}\right)\right\}\,\xi^0\eqno(2.29)$$
so that
$$\lim_{R\to\infty}\hat\xi^{[0}k^{1]}=-{1\over2}\mu\xi^0\,.\eqno(2.30)$$

The conserved charge (2.8) is therefore
$$Q_{[1]}=-{\pi\over\kappa}(\mu\xi^0+2a\xi^2)\eqno(2.31)$$
where, see (2.16), $\kappa=-{8\pi G_3\over c^4}$.

For $\xi^A=\xi^A_{(T)}=(1,0,0)$, and denoting in that case $Q_{[1]}\equiv M_{[1]}\,c^2$, we therefore have, since ${\cal V}_1=2\pi$, that the ``mass parameter" $\mu$ is related
to the  conserved mass $M_{[1]}$ by
$$\mu={8G_3M_{[1]}\over c^2}\,.\eqno(2.32)$$
Comparing this expression with the general  expression (2.22) for $\mu$ in terms of the gravitational mass $M_*$, we find once again that the
conserved mass $M_{[1]}$ and the gravitational mass $M_*$ are equal.

For $\xi^A=\xi^A_{(\phi)}=(0,0,1)$, and denoting in that case $Q_{[1]}\equiv cJ_{[1]}$, formula (2.31) relates the ``angular momentum parameter" $a$ to the conserved angular
momentum $J$ by
$$a={4G_3\over c^3}J_{[1]}\,.\eqno(2.33)$$
Now, in four dimensions, the asymptotically flat form of the metric coefficient $g_{0\phi}$ is (see e.g.~\cite{Str91})~: $g_{0\phi}\to{2G_4J_*\over c^3 R}\,\sin^2\theta$, 
where
$J_*$ is the ``gravitational" angular momentum as measured by, say, gyroscopes. That formula can be generalized to $D$ dimensions as
$$g_{0\phi}\to{8\pi\over{\cal V}_{(D-2)}}\,{G_DJ_*\over c^3 R^{D-3}}\, \Omega_{(D-2)}\eqno(2.34)$$
where, using the angular coordinates of footnote (3), $\Omega_1=1, \Omega_2=\sin^2\theta, \Omega_3=\sin^2\theta\sin^2\chi$, etc, so that
$$g_{0\phi}\to {4G_3J_*\over c^3}\qquad\hbox{for}\qquad D=3\,.\eqno(2.35)$$
Comparing this expression with (2.24) we have
$$a={4G_3\over c^3}J_*\eqno(2.36)$$
and hence, from (2.33), the ``gravitational" angular momentum $J_*$ and the KBL conserved one $J_{[1]}$ are equal, as they must.
Rotating, asymptotically AdS spacetimes have been studied, and their
conserved angular momentum obtained, in, e.g.~\cite{HawHunTay98,AroConOle99,BisMuk03}. 
In~\cite{Sil02} the KBL superpotential was used to compute the
conserved angular momentum of the four dimensional AdS Kerr spacetime.
We compute here the angular momentum of the BTZ solution.

\newpage

\noindent {\csc III. Einstein Gauss-Bonnet theory}
\medskip

In Einstein Gauss-Bonnet theory, gravity is described by the sum of the  Hilbert Lagrangian $L_{[1]}$ and the Lanczos-Lovelock 
Lagrangian~\cite{Lan32}~\cite{Lov71}
$L_{(2)}$
$$L_{[2]}=L_{[1]}+\alpha L_{(2)}\qquad{\rm with}\qquad L_{(2)}\equiv R_{ABCD}R^{ABCD}-4r_{AB}r^{AB}+s^2\eqno(3.1)$$
 where $\alpha$ is some coupling constant. It will be useful to write $L_{(2)}$   as follows~:
$$L_{(2)}=R^{ABCD}P_{ABCD}~~~{\rm where}~~~ P_{ABCD}\equiv R_{ABCD} -2r_{A[C}g_{D]B}+2r_{B[C}g_{D]A}+s\, g_{A[C}g_{D]B}\,.\eqno(3.2)$$
$P_{ABCD}$  has the symmetries of $R_{ABCD}$~\cite{Dav02} and is divergenceless over the indice  $C$ (or $D$) but not over $A$ (or $B$). The variational derivative of (3.1)
with respect to the metric is
$$\sigma^{[2]}_{AB}=\sigma^{[1]}_{AB}+\alpha\,\sigma^{(2)}_{AB}$$
with $\sigma^{[1]}_{AB}$ given in (2.2) and where
$$\sigma^{(2)}_{AB}\equiv 2 R_A^{\ CDE}P_{BCDE} -{1\over2}g_{AB}R^{CDEF}P_{ CDEF}\,.\eqno(3.3)$$  
The vector $V^A$ defined in (1.3) is here of the form
$$V_{(2)}^A=\alpha_{(2)}^{A(BC)}\delta g_{BC}+\beta^{A(BC)}_{(2)~~D}\,\delta\Gamma^D_{BC}\eqno(3.4)$$
 with, see e.g. \cite{DerMad03},
$$\alpha_{(2)}^{A(BC)}=4\alpha\, D^{(B}\sigma_{(1)}^{C)A} ~~~{\rm and}~~~\beta^{A(BC)D}_{(2) }=-4\alpha\left ( P^{A(BC)D}-Q^{A(BC)D}  \right ) \eqno(3.5)$$
in which $Q^{ABCD}$ is antisymmetric in $AB$ but symmetrical in $CD$~:
$$Q^{ABCD}=2(r^{C[A}g^{B]D}+ r^{D[A}g^{B]C})\,.\eqno(3.6)$$
Solving equations (1.13)  yields
$${\cal A}^{[AB]C}_{(2)}=0,~~~{\rm and}~~~{\cal B}_{(2)}^{[AB]CD}=4\alpha\left(P^{ABCD}-Q^{ABCD}\right)\,,\eqno(3.7)$$
so that the first superpotential (1.12) associated with $L_{[2]}=L_{[1]}+\alpha L_{(2)}$ is
$$j^{[AB]}_{[2]a}= 2D^{[A}\xi^{B]}+4\alpha\left[P^{ABCD}D_{[C}\xi_{D]}-Q^{ABCD}D_{(C}\xi_{D)}\right]\,.\eqno(3.8)$$ 
If $\xi^A$ is a Killing vector this expression simplifies since, then~: $ D_{(C}\xi_{D)}=0$.

\bigskip
 {\bf a. A proposal for the Einstein Gauss-Bonnet vector $k^A$}
\medskip

In view of (3.4), and in keeping with the choice made for $k_{[1]}^A$ in (2.5)  in Einstein's theory, we shall choose $k_{(2)}^A$ as~:
$$k_{(2)}^A=-\beta^{A(BC)}_{(2)\ \ \  D}\,\Delta^D_{BC}={\cal B}_{(2)\ \ \ D}^{[AB]C}\,\Delta^D_{BC}=4\alpha \left(P^{ABC}_{\ \ \ \ D}-Q^{ABC}_{\ \ \ \ D} 
\right)\Delta^D_{BC}=$$
$$=4\alpha\left(R^{ABC}_{\ \ \ \ D}+4r^{C[B}\delta^{A]}_D-s\,g^{C[B}\delta^{A]}_D\right)\, \Delta^D_{BC}=\eqno(3.9)$$
$$=4\alpha\left[R^{ABC}_{\ \ \ \
D}\Delta^D_{BC}-2\left(r^{AB}\Delta^C_{BC}-r^{BC}\Delta^A_{BC}\right)+{1\over2}\,s\,\left(g^{AB}\Delta^C_{BC}-g^{BC}\Delta^A_{BC}\right)\right]\,.$$
(See Section IV for further comments on that choice).
Thus the second superpotential (1.12) is chosen to be
$$j^{[AB]}_{[2]b}=2\xi^{[A} k^{B]}_{[2]}\eqno(3.10)$$
with\footnote{Note that the action (1.17) with
$k^A$ given by (3.11) is {\it not} the action proposed by Myers \cite{Mye87}, Davis \cite{Dav02} and Gravanis-Willison \cite{GraWil02}.} 
$$k^A_{[2]}\equiv k_{[1]}^A+k_{(2)}^A=2g^{C[A}\Delta^{B]}_{BC}+4\alpha \left(P^{ABCD}-Q^{ABCD}  \right)\Delta^D_{BC}\,.\eqno(3.11)$$

\bigskip {\bf b. The mass of a static, spherically symmetric and
asymptotically anti-de Sitter spacetime}
\medskip

 Consider again the class of static, spherically symmetric $D$-dimensional spacetimes whose metrics, outside matter, can be written as (2.9)~:
$$ds^2=-f\, c^2\,dT^2+{dR^2\over f}+R^2d\Omega^2_{(D-2)}\,.$$
 When gravity is described by Einstein Gauss-Bonnet theory with $\Lambda<0$,   $f$ is given (outside matter if any) by~\cite{BouDes85} 
$$f(R)=1+{R^2\over2\tilde\alpha}-{R^2\over2\tilde\alpha}\sqrt{1-{4\tilde\alpha\over\ell^2}+{4\mu\tilde\alpha\over R^{D-1}}}\eqno(3.12)$$
 where $\ell^2=-{(D-1)(D-2)\over2\Lambda}$, where $\tilde\alpha=(D-3)(D-4)\alpha$ and where $\mu$ is the ``mass" parameter, that we want to relate to the conserved mass
of  spacetime as defined above.

We choose again the background  spacetime $\bar{\cal M}$ to be the $\mu=0$ anti-de Sitter spacetime (or Minkowski if the cosmological constant is set to zero), that is 
$$\bar f=1+{R^2\over {\cal L}^2}\qquad\hbox{with the length scale ${\cal L}$ given
by}\qquad{1\over{\cal L}^2}={1\over2\tilde\alpha}\left(1-\sqrt{1-{4\tilde\alpha\over\ell^2}}\right)\eqno(3.13)$$ 
so that
$$ \Delta f\equiv f-\bar f\to-{\tilde\mu\over R^{D-3}}\qquad\hbox{with}\qquad\tilde\mu={\mu\over\sqrt{1-{4\tilde\alpha\over\ell^2}}}\qquad\hbox{when}\qquad
{R\to\infty}\,.\eqno(3.14)$$

Finally, we choose again $\xi^A$ to be the time translation Killing vector normalized as $\xi^A=(1,0,...,0)$, and we relate as before the coupling constant $\kappa$ to the
$D$-dimensional Newton constant by $\kappa=-{8\pi G_D\over c^4}$.

The conserved mass $M_{[2]}$ of spacetime (with respect to anti-de Sitter) is therefore defined as
$$M_{[2]}=M_{[2]a}+M_{[2]b}\qquad\hbox{with}\qquad  \kappa M_{[2]b}\,c^2=\int_{\cal S}\!d^{D-2}x\ \hat\xi^{[0}k^{1]}_{[2]}\eqno(3.15)$$
 and
$$ M_{[2]a}\,c^2=\int_{\cal S}\!d^{D-2}x\ \hat J^{[01]}_{[2]a}\quad\hbox{with}\quad 2\kappa\hat J^{[AB]}_{[2]a}\equiv\hat j^{[AB]}_{[2]a}-\overline{\hat
j^{[AB]}_{[2]a}}\eqno(3.16)$$
 where $j_{[2]a}^{[AB]}$ is given by (3.8).

The steps to get $M_{[2]a}$ are~: 
$$j_{[2]a}^{[01]}=2D^{[0}\xi^{1]}+4\alpha P^{01CD}D_{[C}\xi_{D]}=-f'+4\alpha f'P_{0101}\,,$$
Now, from (3.2)~:
$$P_{0101}=R_{0101}+fr_{11}-{1\over f}\,r_{00}-{1\over2}s\,,$$ 
that is, using the explicit expression of the Riemann tensor for the metric (2.9)~:\footnote{that is~: $R_{0101}={1\over2}f''$, $R_{0i0j}={1\over2}Rff'\tilde g_{ij}$,
$R_{1i1j}=-{1\over2}R{f'\over f}\tilde g_{ij}$, $R_{ijkl}=(1-f)R^2(\tilde g_{ik}\tilde g_{jl}-\tilde g_{jk}\tilde g_{il})$ where $\tilde g_{ij}$ are the coefficients of the metric
on the unit $(D-2)$-sphere}
$$P_{0101}=-{(1-f)\over2R^2}(D-2)(D-3)$$ and hence~:
$$ j_{[2]a}^{[01]}=-f'\left[1+2\alpha {(1-f)\over R^2}(D-2)(D-3)\right]$$ so that, finally~:
$$\kappa M_{[2]a}\, c^2=-{1\over2}{\cal V}_{(D-2)}\lim_{R\to\infty}R^{D-2}\Delta f'+\hskip 9cm$$
$$+\alpha\lim_{R\to\infty}{\cal V}_{(D-2)}R^{D-4}(D-2)(D-3)[\Delta f(\bar f'+\Delta f')-\Delta f'(1-\bar f)]\,.\eqno(3.17)$$
 We now inject in this expression the solution (3.12-14) and get
$$\kappa M_{[2]a}\, c^2=-\tilde\mu {\cal V}_{(D-2)}{(D-3)\over2}\left[1-{(D-2)(D-5)\over(D-3)(D-4)}\left(1-\sqrt{1-{4\tilde\alpha\over\ell^2}}\right)\right]\,.\eqno(3.18)$$
(Note that when the cosmological constant is zero, that is $\ell\to\infty$ then $\mu$ replaces $\tilde\mu$ and $\bar f=1$ and we get for $M_{[2]a}$ the same result as in pure
Einstein's theory, that is (2.18).)

As for the vector $k^A_{(2)}$ it is given by (3.9). It is easy to compute the limit of its $R$-component $k^1_{(2)}$ when $R\to\infty$. Indeed one can replace the $P^{ABCD}$ and
$Q^{ABCD}$  tensors in (3.9) by their asymptotic, anti-de Sitter expressions. One also uses the fact that $\lim_{R\to\infty}\Delta^1_{BC}g^{BC}=-\Delta f'-{(D-2)\over R}\Delta
f$ and $\lim_{R\to\infty}\Delta^C_{1C}=0$; one then gets
$$\lim_{R\to\infty}k^1_{[2]}=\left[1-{2\alpha\over{\cal L}^2}(D-2)(D-3)\right]\lim_{R\to\infty}\left[\Delta f'+{(D-2)\over R}\Delta f\right]\,.$$
 If we now inject the expression (3.14) for $\Delta f$, we obtain, see (3.15)
$$\kappa M_{[2]b}\, c^2=-{1\over2}\tilde\mu{\cal V}_{(D-2)}\left[1-{(D-2)\over (D-4)}\left(1-\sqrt{1-{4\tilde\alpha\over\ell^2}}\right)\right]\,.\eqno(3.19)$$

Adding the two contributions, $M_{[2]a}$ as given by (3.18) and $M_{[2]b}$ as given by (3.19),we obtain that the mass parameter $\mu$ of a static, spherically symmetric
asymptotically  anti-de Sitter spacetime is related to the conserved mass $M_{[2]}=M_{[2]a}+M_{[2]b}$ by
$$\mu={16\pi G_D\over c^2{\cal V}_{(D-2)}}{M_{[2]}\over(D-2)}$$
 and, hence, see (2.22), the gravitational mass $M_*$ and the KBL conserved mass $M_{[2]}$ are equal, a result which justifies the choice (3.11) for the vector $k^A_{[2]}$.  In this
particular example of AdS Einstein Gauss-Bonnet static spherically symmetric spacetime  there is perfect agreement between various definitions of  mass-energy, for example,
that of~\cite{DesTek02}~\cite{JacMye93}~\cite{BanTeiZan93}~\cite{LouSimWin96} or that
of~\cite{Cai01}~\cite{NojOdiOgu01}~\cite{CveNojOdi01}~\cite{ChoNeu02}~\cite{Neu03}.\footnote{In
~\cite{Cai01}~\cite{NojOdiOgu01}~\cite{CveNojOdi01}~\cite{ChoNeu02}~\cite{Neu03} the result is obtained by means of the Euclidean path integral method, using a different
boundary term for the action (but the boundary term does not contribute to the action on shell) and a different regularization procedure. See~\cite{DerOgu03} for a detailed
comparison.}

\bigskip
\noindent {\csc IV. Critical comments}
\medskip

Let us point out two shortcomings of the approach used in this paper. One is the choice (2.12) (the same as in~\cite{DesTek02}) made for the
normalisation of the Killing vector when spacetime is not asymptotically flat~: as we have already mentionned there is no profound reason (see e.g.~\cite{JacMye93}) to choose
that (asymptotically divergent) norm ($g_{AB}\xi^A\xi^B\to- R^2/\ell^2$ when $R\to\infty$).

A more severe criticism perhaps concerns the choice (3.9) we made for the vector $k^A_{(2)}$. Contrarily to $k^A_{[1]}$, which cancels all the derivatives of the metric
variations in the boundary term of the variation of the Einstein-Hilbert action (see (1.18)), the vector $k^A_{(2)}$ does {\it not} cancel all the normal derivatives of the
variation of the Einstein-Hilbert Gauss-Bonnet action. In order to fully justify our choice (which yields the right mass) it remains to see whether extra terms can be added to
cure that and to check that they do not ``spoil" our equality of the conserved mass $M_{[2]}$ and the gravitational mass
$M_*$. Another test that the vector $k^A_{(2)}$ should pass is to check whether or
not it gives the right angular momentum: this check should not be too
difficult to perform as some rotating solutions are already known in
Einstein Gauss-Bonnet gravity \cite{Deh}. Finally the calculation of the
Bondi mass in Einstein Gauss-Bonnet theory remains to be done. We leave
these questions to future work.

\bigskip
\noindent
{\bf Acknowledgements~: } N.D. thanks the Kyoto Yukawa Institute for Theoretical Physics as well as the Cambridge Centre for Mathematical Sciences, and J.K. thanks
the Cambridge Institute of Astronomy for their hospitality and financial support during the time this work was performed. N.D. thanks Ted Jacobson for fruitful discussions. S.O. thanks
Kei-ichi Maeda and Ishwaree P. Neupane for useful discussions.  
The research of S.O. is supported by the Japan Society for the Promotion of Science.

\bigskip
\newpage
\providecommand{\href}[2]{#2}\begingroup\raggedright\endgroup

\end{document}